\documentclass[9pt, conference]{IEEEtran}

\IEEEoverridecommandlockouts

\usepackage{cite}
\usepackage{amsmath,amssymb,amsfonts}

\usepackage{graphicx}
\usepackage{textcomp}
\usepackage{xcolor}
\usepackage{amssymb}
\usepackage{pifont}
\usepackage{algorithmic}
\usepackage[ruled,vlined,linesnumbered]{algorithm2e}

\DontPrintSemicolon
\SetNoFillComment
\newcommand{\cmark}{\ding{51}}%
\newcommand{\xmark}{\ding{55}}%

\let\oldnl\nl
\newcommand{\nlnonumber}{\renewcommand{\nl}{\let\nl\oldnl}}
\SetCommentSty{mycommfont}
\usepackage{threeparttable}
\usepackage{booktabs}
\def\BibTeX{{\rm B\kern-.05em{\sc i\kern-.025em b}\kern-.08em
    T\kern-.1667em\lower.7ex\hbox{E}\kern-.125emX}}
\begin{document}

\title{Data Efficient Child-Adult Speaker Diarization with Simulated Conversations \thanks{Thank you to Apple and Simons Foundation for the funding.}}

\author{Anfeng Xu$^{\star}$, Tiantian Feng$^{\star}$, 
 Helen Tager-Flusberg$^{\dagger}$, Catherine Lord$^{\ddagger}$, Shrikanth Narayanan$^{\star}$ \\ \\

\IEEEauthorblockA{
University of Southern California,
Los Angeles, USA$^{\star}$}

\IEEEauthorblockA{
Boston University, Boston, USA$^{\dagger}$}
\IEEEauthorblockA{
University of California, Los Angeles, Los Angeles, USA$^{\ddagger}$}
}

\maketitle

\begin{abstract}
Automating child speech analysis is crucial for applications such as neurocognitive assessments. Speaker diarization, which identifies ``who spoke when'', is an essential component of the automated analysis. However, publicly available child-adult speaker diarization solutions are scarce due to privacy concerns and a lack of annotated datasets, while manually annotating data for each scenario is both time-consuming and costly. To overcome these challenges, we propose a data-efficient solution by creating simulated child-adult conversations using AudioSet. We then train a Whisper Encoder-based model, achieving strong zero-shot performance on child-adult speaker diarization using real datasets. The model performance improves substantially when fine-tuned with only 30 minutes of real train data, with LoRA further improving the transfer learning performance. The source code and the child-adult speaker diarization model trained on simulated conversations are publicly available.
\end{abstract}


\section{Introduction}
\label{sec:intro}
\begin{figure}[t]
  \centering
  \includegraphics[width=0.96\linewidth]{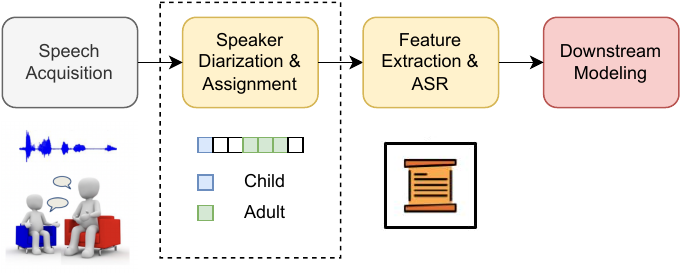}
  \vspace{-3mm}
  \caption{Spoken language assessment pipeline.}
  \label{fig:pipeline}
  \vspace{-3mm}
\end{figure}

A wide range of critical child-centered applications, including neurocognitive assessments, psychotherapy interventions, personalized learning, and child forensic interviews, depend on direct, often dyadic interactions between the child and an adult, such as a clinician, caregiver, or teacher.
For example, the Autism Spectrum Disorder (ASD) assessment relies on child-adult dyadic interaction sessions such as ADOS-Mod2~\cite{mccrimmon2014test} and ELSA \cite{barokova2021eliciting}.
Automating the analysis and assessment of children's speech and language can substantially reduce the associated costs and expedite the diagnostic process.
A crucial step towards automated child speech analysis is speaker diarization, which answers the question of ``who spoke when'', as illustrated in Fig~\ref{fig:pipeline}.
For instance, an earlier study \cite{xu2023understanding} has established that the amount of intelligible speech produced by a child, which can be derived directly through child-adult speaker diarization, indicates the child’s developmental language proficiency.

Recent advancements in deep learning have considerably accelerated the development of robust speaker diarization systems \cite{park2022review}. 
The most critical task in the child-adult speaker diarization setup is recognizing the child-adult role. 
Researchers have proposed to formulate the problem as a speaker classification problem, directly assigning \textit{child} and \textit{adult} speaker labels, such as Kothalkar et. al~\cite{kothalkar2024child} using CNNs and Lavechin et. al\cite{lavechin2020open} using LSTM to perform segment-level or frame-level child-adult classification. 
Recently, pre-trained speech models have shown promise in this domain for their advanced capabilities in distinguishing child voices from adult voices. 
Earlier efforts showed that pre-trained models such as WavLM~\cite{chen2022wavlm}, wav2vec 2.0~\cite{baevski2020wav2vec}, and Whisper Encoder~\cite{radford2023robust} show remarkable performance on the utterance-level child-adult speaker classification task~\cite{lahiri2023robust, xu2024audio}. 
Additionally, Li et. al~\cite{li2023towards} used wav2vec 2.0 to perform frame-level speaker classification for infant-adult speaker diarization. 
Most recently, \cite{xu2024exploring} investigated various speech foundation models, pre-trained on vast speech data from the internet, on the child-adult speaker diarization task, showing a substantially lower diarization error rate (DER) compared to previous methods.

Despite the aforementioned endeavors, public access to child-adult speaker classification or diarization models is still limited, mainly due to insufficient public child-adult interaction datasets with appropriate annotations for training speaker diarization models. 
For example, some popular datasets in this domain such as the CHILDES corpora from TalkBank database~\cite{macwhinney2007talkbank} lack fine-grained speaker timestamp information while the relevant data from DIHARD challenge~\cite{ryant2019second} focus on infants or teenage students.
Additionally, most previous works have focused on highly sensitive, private datasets, for which model release would raise privacy concerns.
Moreover, these works assume the availability of a sufficient amount of child-adult interaction data with adequate annotations, which are both costly and time-consuming to acquire in practical scenarios.

To address these challenges, we construct simulated child-adult conversations from AudioSet \cite{gemmeke2017audio} and subsequently train speaker diarization models based on the Whisper Encoder. The main contributions and findings in this work are summarized below:

\begin{itemize}

    \item 
    We proposed a comprehensive algorithm to simulate child-adult conversations using Audioset. This algorithm accounts for variables and conditions such as probabilities of overlapping speech, pauses or silences between turns, and background noises. The Whisper Encoder trained on the simulated conversations achieves remarkable zero-shot child-adult speaker diarization.
    \item 
    Our extensive fine-tuning experiments show outstanding transfer learning capabilities when fine-tuned with a small amount of real child-adult speech data. We also find that Low-Rank Adaptation (LoRA)~\cite{hu2021lora} further improves the transfer learning performance. 
    
    \item 
    The code for constructing simulated dialogues is shared with the public to facilitate child-adult speaker diarization research in the community.
    Moreover, we publicly release a child-adult speaker diarization model using a Whisper-base Encoder trained on 50k simulated conversations. 
    The code and model weights are available at \textit{https://github.com/usc-sail/child-adult-diarization}.

\end{itemize}

\section{Background}
\subsection{Speaker diarization}
Traditional speaker diarization typically starts with Voice Activity Detection (VAD) to extract regions containing human speech. 
Then, speaker embeddings, such as x-vectors~\cite{snyder2018x}, are computed from these speech segments and clustered to assign speaker labels. The current state-of-the-art speaker diarization algorithm with clustering is VBx~\cite{landini2022bayesian}, which employs a hidden Markov model to refine speaker embeddings and re-segment speaker regions based on initial clusters generated through Agglomerative Hierarchical Clustering (AHC) on x-vectors.
Recently, to address overlapping speech, researchers introduced End-to-End Neural Diarization (EEND)~\cite{fujita2019end}, which uses neural networks to treat diarization as a classification problem. 
To improve EEND, a hybrid model named EEND-VC~\cite{kinoshita2021integrating, kinoshita2021advances} was developed, using EEND on speech sub-segments and then using clustering to combine the results from sub-segments for the final speaker labeling. 
In this study, we employ VBx~\cite{landini2022bayesian} for clustering-based diarization and PyAnnote \cite{plaquet2023powerset} Diarization for EEND-VC diarization, as the baselines for comparison. However, it is worth noting that these are general-purpose diarization models that can be used for any number of speakers without any prior speaker role presumptions (e.g., child-adult, customer-operator). In contrast, our objective focuses on a specialized case of the speaker diarization problem, where we aim to develop a model specifically tailored to distinguish between child and adult speakers in dyadic settings.
\subsection{Simulated conversation}
When training EEND models, researchers have relied on using simulated conversations as the initial training data to overcome the challenge of data scarcity in speaker diarization. Early studies in EEND have used simulated mixtures, which are created by simply summing up the single-speaker recordings \cite{fujita2019end}. Recent works have proposed to generate more naturalistic simulated conversations by considering the turn-taking, pause, and overlaps to improve the performance of speaker diarization models \cite{landini2022simulated, yamashita2022improving}. In this study, we propose a child-adult conversation simulation algorithm inspired by these works. 

\section{Methods}
\subsection{Dataset for Simulated Child-Adult Conversation}

We collect speech segments from AudioSet \cite{gemmeke2017audio}, which is a large-scale dataset consisting of 10-second audio files from YouTube videos with manually annotated audio event labels. Specifically, we use audio files labeled as \textit{male speech, man speaking} and \textit{Female speech, woman speaking} as the adult male and female speech data and audio files labeled as \textit{child speech, kid speaking} as the child speech data, all from the \textit{unbalanced\_train} set. 
In total, we have obtained 14,524, 6,904, and 8,515 audio files for adult male, adult female, and child speech, respectively.

Since these 10-second segments contain non-speech segments, such as silence or background noise, we use a pre-trained internal speaker diarization model from \cite{xu2024exploring} to extract speech segments. 
Given an adult speaker ${s_a}$ or a child speaker ${s_c}$, we denote the set of child and adult speech segments extracted from a 10-segment audio file as $U_{s_c}$ and $U_{s_a}$ respectively. 9,146, 4,749, and 3,826 audio files are used after the pre-processing for adult male, adult female, and child speakers.
On average, we extract 4.16, 4.39, and 3.46 speech segments for each audio file with an average duration of 1.53, 1.49, and 1.00 seconds for adult male, adult female, and child speech respectively.
We have also considered generic VAD models such as \cite{Bredin2021, Silero-VAD} and speech enhancement models such as \cite{defossez2020real} for the pre-processing. However, our internal experiments reveal that these models often incorrectly filter out child speech since they are trained mainly on adult speech.


\subsection{Child-Adult Conversation Simulation Algorithm}

\SetKwComment{Comment}{  $\triangleright$\ }{\!}
\begin{algorithm}[t]
 \label{alg:algorithm1}
\footnotesize
\SetKwInput{Input}{Input}
\SetKwInput{Output}{Output}
\SetKwInput{Init}{Initialization}
\Input{$p_o, p_c, p_{st}, \beta_{intra}$, $\beta_{inter}$ \Comment*[r]{Hyper-parameters}\\
       \nlnonumber $S_a$, $S_c$ \Comment*[r]{Set of adult \& child speakers}\\ 
       \nlnonumber $\mathcal{U}_a= \{U_{s_a}\}_{s_a \in S_a}$ \Comment*[r]{Set of adult utterances} \\
       \nlnonumber $\mathcal{U}_c= \{U_{s_c}\}_{s_c \in S_c}$ \Comment*[r]{Set of child utterances}}

\Output{y \Comment*[r]{10s audio for output}}
\SetAlgoLined
\Init{y = []\Comment*[r]{initialize empty output} \\
\nlnonumber Sample $s_a$ and $s_c$ from $S_a$ and $S_c$ with replacement}
 
sample $p$ from Uniform($[0,1]$) \Comment*[r]{utterance at start}
 \If{p $<$ $p_{st}$}{
 sample $p$ from Uniform($[0,1]$) \Comment*[r]{child or adult}
    \eIf{$p < p_{c}$}{
        sample $u$ from $U_{s_c}$ 
    }{
        sample $u$ from $U_{s_a}$ 
  }
     $y \gets y \oplus u_{\text{Uniform}[0, end]:end}$  
 }
 prepare silence $sil$ of length sampled from $Exp(\beta_{intra})$ \\
 $y \gets y \oplus sil$ \\
 \While{length(y) $<$ 10s}{
    sample $p$ from Uniform($[0,1]$) \Comment*[r]{child or adult}
    \eIf{$p < p_{c}$}{
        sample $u$ from $U_{s_c}$ 
    }{
        sample $u$ from $U_{s_a}$ 
  }
    \eIf{last utterance from the same speaker}{
        prepare  $sil$ of length sampled from $Exp(\beta_{intra})$ \\
         $y \gets y \oplus u \oplus sil$ \\
    }{
        sample $p$ from Uniform($[0,1]$) \Comment*[r]{overlap or not}
        \eIf{$p < p_{o}$}{
            add $u$ to $y$ starting at random position of last utterance
        }{
            prepare $sil$ of length sampled from $Exp(\beta_{inter})$ \\
            $y \gets y \oplus u \oplus sil$ \\
        }
    }
 }
 return $y_{0:10s}$ + noise

 \caption{child-adult conversation simulation}
\end{algorithm}
Motivated by the works in \cite{landini2022simulated, yamashita2022improving}, we design the simulated conversation by incrementally appending speech segments with the silence duration modeled by an exponential distribution, as detailed in Algorithm~\ref{alg:algorithm1}. 
$p_o$, $p_c$, and $p_{st}$ represent the probabilities of overlap, child speech, and the simulated conversation starting with speech, respectively. $\beta_{intra}$ and $\beta_{inter}$ represent the scale parameters of the exponential distribution for intra-speaker and inter-speaker silences, respectively. 
We sample a speech segment $u$ from $U_{s_c}$ or $U_{s_a}$ without replacement, and if either set is exhausted, it is replenished from the original. The noise is sampled from background noises from the Musan corpus~\cite{snyder2015musan}, with SNR randomly selected from 5dB, 10dB, 15dB, and 20dB. 

In addition, we set 20\% of the simulated conversation to contain no speech segments to prevent the model from outputting speaker predictions in the absence of actual speech. We select females as the adult speech 85\% of the time since we empirically found that distinguishing adult female voices from child voices is more challenging than distinguishing adult male voices from child voices. 
We use 0.1, 0.4, 0.5, 1, and 0.8 as the hyper-parameters for $p_o$, $p_c$, $p_{st},\beta_{intra}$, and $\beta_{inter}$. The hyper-parameters are selected based on the statistics from Remote-NLS.
We set the simulated conversation length to 10 seconds, matching the duration of audio files in Audioset, though the length can be adjusted as needed.

\subsection{Modeling Architecture with a Whisper Encoder}
We formulate the child-adult speaker diarization problem as a frame-level classification problem with a Whisper Encoder, which has shown remarkable performances in a prior work~\cite{xu2024exploring}.
Given an audio input $X = [x_1, ..., x_T]$, the goal is to predict the speech label $y_t \in \{s, c, a, o\}$ for $x_t$ at each timestamp $t$, where $s, c, a, $ and $o$ represent silence/noise, child speech, adult speech, and overlapping speech, respectively.
In our work, $T = 500$ since the input $X$ is 10 seconds long and the step size of a Whisper Encoder is 20 ms.

\begin{figure}[t]
  \centering
  \includegraphics[width=0.9\linewidth]{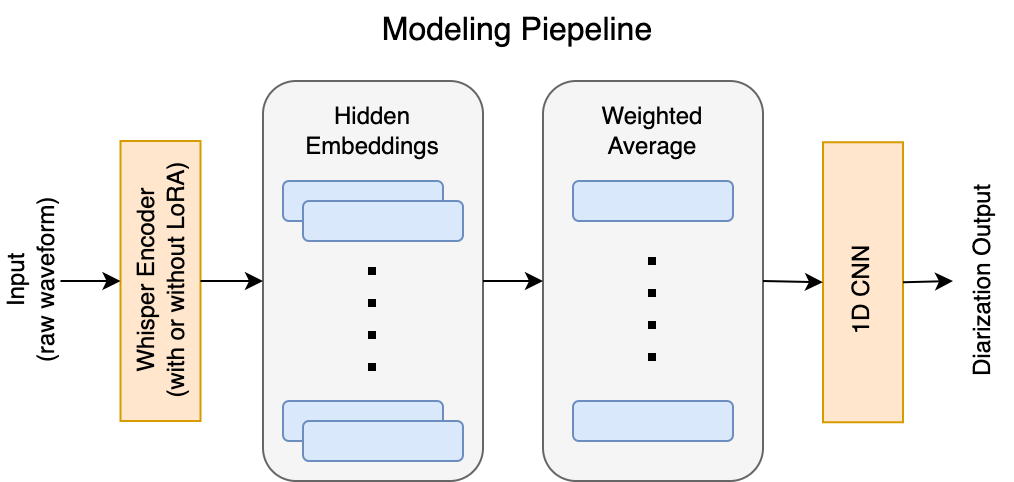}
  \caption{Speaker diarization pipeline with Whisper Encoder.}
  \label{fig:modeling_pipeline}
\end{figure}

The modeling pipeline is illustrated in Fig~\ref{fig:modeling_pipeline}. Here, we freeze the Whisper Encoder layers. In addition, we apply Low-Rank Adaptation (LoRA)~\cite{hu2021lora}, which fine-tunes low-rank matrices with a small number of parameters to approximate model updates for transformers. In our experiments, we apply LoRA at the feed-forward layers of transformers inside the Whisper Encoder.
The input waveform is first passed to a Whisper Encoder and the hidden embeddings are extracted. 
We take the weighted average of the hidden embeddings, where the weights are learnable. 
Then, three 1D CNNs each with a channel size of 256 are applied. When using LoRA, two 1D CNNs are used instead. 
Each CNN has a kernel size of 1 and a dropout probability of 0.2, followed by a ReLU activation function.
Finally, a 1D CNN with a kernel size of 1 and channel width of 4 is applied to get the final predictions.

\section{Experiments}
\subsection{Datasets}
We consider two datasets containing child-adult dyadic sessions in clinical contexts to evaluate the effectiveness of simulated conversations for child-adult speaker diarization. We adhere to the data usage terms specified in the IRB and Data Use Agreements (DUAs) provided by the original data owners.

\textbf{Remote-NLS} includes 73 Zoom recordings, each consisting of 15 minutes of child-parent dyadic interactions that would elicit the child's spontaneous speech in a natural environment, as described in \cite{butler2022remote}. All the children are diagnosed with ASD and the majority of them are minimally verbal. The children are between 4 and 7 years old, and 18 are females. The audio files are annotated for speaker diarization experiments purposes as described in \cite{xu2024exploring}. We use this dataset for both zero-shot and transfer learning experiments for child-adult speaker diarization. 

\textbf{ADOS-Mod3} consists of 335 audio recordings from 171 children, using the ``Social Difficulties'' and ``Annoyance and Emotional'' sub-tasks in the ADOS-2 protocol for ASD diagnosis, as described in~\cite{lahiri2022interpersonal}. 
The majority of the children are verbally fluent. The children are aged between 2 to 13 years old, and 45 are females. Around half of them are diagnosed as having ASD, while most others are diagnosed with other disorders such as ADHD and mental or language disorders.
The audio files are annotated at utterance-level and the unintelligible speech segments are ignored. Thus, we use this dataset only for the zero-shot evaluation.
Additionally, we separate the dataset between 66 children 2 to 7 years (ADOS-young) and 105 children 8 to 13 years (ADOS-old) for the evaluation.

\subsection{Baseline Speaker Diarization Models}
We apply PyAnnote Diarization~\cite{plaquet2023powerset}, as well as VBx~\cite{landini2022bayesian} with PyAnnote VAD~\cite{Bredin2021}, using the code and pre-trained models provided by the authors. For PyAnnote Diarization, we set the minimum and maximum number of speakers to one and two.
For VBx, we use FA = 0.3, FB = 12, and loop$_P$ = 0.2 as the hyper-parameters, determined by grid search.
We also report the results after fine-tuning the audio segmentation models in PyAnnote Diarization and PyAnnote VAD (for VBx) using the Remote-NLS dataset.
We train for 15 epochs using Adam optimizer with a learning rate of $1e^{-3}$, batch size of 32, and Cross-Entropy Loss.

\subsection{Training Details for Whisper Encoders}
We use the pre-trained Whisper-base Encoder available on HuggingFace~\cite{wolf2019huggingface}. We use the Adam optimizer with Cross-Entropy Loss, a learning rate of 5e-4, and a weight decay of 1e-4, for up to 20 epochs. The same hyper-parameters, determined empirically, are used for both training on simulated datasets and fine-tuning with the Remote-NLS dataset. 

For the fine-tuning experiments with Remote-NLS, we conduct 5-fold cross-validation at the session level to prevent any overlap of speakers between the training and test sets. During training, 25\% of the sessions are reserved for validation. Training samples are drawn with a 50\% overlap between adjacent audio windows, while test samples are processed with non-overlapping sliding windows, where each window is 10 seconds. In the case of LoRA fine-tuning, we set the low-rank dimension as 8. 
\subsection{Evaluation}
We report the diarization error rate (DER), which is the sum of false alarm (FA), missed detection (MD), and speaker confusion (SC) rates, as a percentage using PyAnnote.metrics~\cite{bredin2017pyannote}. We apply a 100 ms forgiveness collar at the boundaries of each reference audio segment and include the overlapping regions. We opt for 100 ms rather than the standard 250 ms because most child audio segments are short. We aggregate the results from all sessions to calculate the evaluation metrics.

\section{Results and analysis}
\subsection{Zero-shot Child-adult Speaker Diarization Results}

The zero-shot speaker diarization results on Remote-NLS using a Whisper-base Encoder (WSP) trained on 50k simulated conversations are reported in Table~\ref{tab:zeroshot-NLS}, along with the baseline diarization results. We see competitive zero-shot results against the baseline in terms of diarization error rates (DER), especially for the speaker confusion (SC) rates. The baseline methods show high missed detection (MD) rates, probably because the models are biased towards adult speech and they miss substantial portions of child speech. In contrast, the Whisper-base Encoder exhibits a high false alarm rate due to overfitting to the noisy child speech segments in the Audioset. As a result, it frequently misclassifies background noise as child speech.

Moreover, the zero-shot results on ADOS-MOD3 are shown in Table~\ref{tab:zeroshot-ados}. Similar to the results from the Remote-NLS dataset, our method substantially improves on the SC metric on younger children between 2 to 7, but shows a notable decline in performance for older children aged 8 to 13. We presume this is because the extracted child speech segments from AudioSet skew towards younger age and distinguishing older children from adults is harder than distinguishing younger children from adults. We observe overall high FA and MD rates, likely due to misaligned annotations for speaker diarization.

Table~\ref{tab:pretrain_data} reports DERs and SC rates with varying numbers of simulated conversations ranging from 5K, 10K, 20K, and 50K using a Whisper-base Encoder with LoRA. We notice that the DERs are comparable even with less number of simulated conversations. This result suggests that scaling the amount of simulated conversation is not likely to improve the speaker diarization performance and that future work may focus on improving the simulated data quality.

\begin{table}[t]

  \caption{Zero-shot Child-adult speaker diarization results on Remote-NLS dataset.}
  \label{tab:zeroshot-NLS}
  \centering
  \vspace{-2.5mm}
  \begin{tabular*}{0.8\linewidth}{l c c c c c c}
    \toprule
    \multicolumn{1}{l}{\textbf{Model}} & \textbf{LoRA} & \textbf{FA} & \textbf{MD} & \textbf{SC} & \textbf{DER} \\
    \cmidrule(lr){1-1} \cmidrule(lr){2-2} \cmidrule(lr){3-6} \cmidrule(lr){6-6} 
    PyAnnote &  N/A & $6.3$ & $12.5$ & $16.1$ & $34.9$ \\
    VBx & N/A & $8.8$ & $15.7$ & $12.2$ & $35.7$ \\
    \cmidrule(lr){1-1} \cmidrule(lr){2-2} \cmidrule(lr){3-6} \cmidrule(lr){6-6} 
    WSP-Base &  \xmark & $15.6$ & $5.0$ & $10.5$ & $\mathbf{31.1}$ \\
    WSP-Base &  \cmark & $18.5$  & $4.3$ & $\mathbf{8.5}$ & $31.3$ \\
    \bottomrule
  \end{tabular*}
\end{table}

\begin{table}[t]

  \caption{Zero-shot Child-adult speaker diarization results on ADOS-Mod3 dataset.}
  \label{tab:zeroshot-ados}
  \centering
  \vspace{-2.5mm}
  \begin{tabular*}{0.9\linewidth}{l c c c c c c}
    \toprule
    \multicolumn{1}{l}{\textbf{Model}} & \textbf{Age} & \textbf{LoRA} & \textbf{FA} & \textbf{MD} & \textbf{SC} & \textbf{DER} \\
    \cmidrule(lr){1-1} \cmidrule(lr){2-2} \cmidrule(lr){3-3} \cmidrule(lr){4-6} \cmidrule(lr){7-7} 
    PyAnnote & 2-7 & N/A & $16.7$ & $19.8$ & $14.4$ & $50.9$ \\
    VBx & 2-7 & N/A & $17.9$  & $23.7$ & $8.8$ & $50.4$ \\
    \cmidrule(lr){1-1} \cmidrule(lr){2-2} 
    \cmidrule(lr){3-3} \cmidrule(lr){4-6} \cmidrule(lr){7-7} 
    WSP-Base &  2-7 & \xmark & $20.7$ & $17.0$ & $\mathbf{5.5}$ & $\mathbf{43.1}$ \\
    WSP-Base &  2-7 & \cmark & $20.5$ & $17.7$ & $5.9$ & $44.1$ \\
    \cmidrule(lr){1-7}
    PyAnnote & 8-13 & N/A & $12.5$ & $20.6$ & $15.2$ & $48.4$ \\
    VBx & 8-13 & N/A & $12.8$ & $25.1$ & $\mathbf{8.9}$ & $\mathbf{46.8}$ \\
    \cmidrule(lr){1-1} \cmidrule(lr){2-2} 
    \cmidrule(lr){3-3} \cmidrule(lr){4-6} \cmidrule(lr){7-7} 
    WSP-Base &  8-13 & \xmark & $15.2$ & $18.7$ & $22.9$ & $56.8$ \\
    WSP-Base &  8-13 & \cmark & $14.9$ & $19.8$ & $25.7$ & $60.4$ \\
    \bottomrule
  \end{tabular*}
\end{table}

\begin{table}[t]
  \caption{Impact of Amount of Simulated Training Data on Zero-shot Child-adult Speaker Diarization Performance: DER (SC).}
  \label{tab:pretrain_data}
  \centering
  \vspace{-2.5mm}
  \begin{tabular*}{\linewidth}{l c c c c}
    \toprule
    \multicolumn{1}{l}{\textbf{Dataset}} &\textbf{5K} &\textbf{10K} & \textbf{20K} & \textbf{50K} \\
    \cmidrule(lr){1-1} \cmidrule(lr){2-5}
    Remote-NLS & $32.0(9.3)$ & $33.3(8.58)$ & $32.2(8.75)$ & $31.3(8.54)$ \\
    ADOS-young & $43.9(5.3)$ & $41.8(3.8)$ & $41.6(3.6)$ & $44.1(5.9)$ \\
    ADOS-old & $59.2(24.0)$ & $55.4(20.7)$ & $54.7(20.0)$ & $60.4(25.7)$ \\
    \bottomrule
  \end{tabular*}
\end{table}


\begin{table}[t]

  \caption{Transfer Learning performance.}
  \label{tab:transfer_learning}
  \centering
  \vspace{-2.5mm}
  \begin{tabular*}{0.94\linewidth}{l c c c c c c c}
    \toprule
    \multicolumn{1}{l}{\textbf{Model}} & \textbf{Pre-trained} &\textbf{LoRA} & \textbf{FA} & \textbf{MD} & \textbf{SC} & \textbf{DER} \\
    \cmidrule(lr){1-1} \cmidrule(lr){2-2} \cmidrule(lr){3-3} \cmidrule(lr){4-6} \cmidrule(lr){7-7} 
    PyAnnote & \xmark & N/A & $2.9$ & $8.0$ & $16.3$ & $27.2$ \\
    VBx & \xmark & N/A & $2.9$  & $9.1$ & $13.9$ & $25.9$ \\
    \cmidrule(lr){1-1} \cmidrule(lr){2-2} \cmidrule(lr){3-3} \cmidrule(lr){4-6} \cmidrule(lr){7-7} 
    
    WSP-Base & \xmark & \xmark & $3.0$ & $8.5$ & $9.9$ & $21.5$ \\
    WSP-Base & \xmark & \cmark & $4.1$ & $6.3$ & $6.3$ & $16.7$\\
    WSP-Base & \cmark & \xmark &$3.2$ & $5.7$ & $10.8$ & $19.6$ \\
    WSP-Base & \cmark & \cmark &$4.7$ & $5.8$ & $\mathbf{5.7}$ & $\mathbf{16.2}$ \\
    \bottomrule
  \end{tabular*}
\vspace{-2.5mm}
\end{table}

\subsection{Transfer Learning Capabilities}
In Table~\ref{tab:transfer_learning}, we report DERs from the fine-tuned Whisper-base Encoder with and without LoRA or pre-training on 50K simulated conversations, along with the baseline fine-tuned speaker diarization results. We can see the advantages of our methods over existing diarization methods, notably for the SC rates. Additionally, adding LoRA substantially improves upon the Whisper-base Encoder performance for the SC rates. We can also see moderate improvements in terms of DER with pre-training on simulated conversations.

\begin{figure}[!t]
  \centering
  \includegraphics[width=8cm, height=5.5cm]{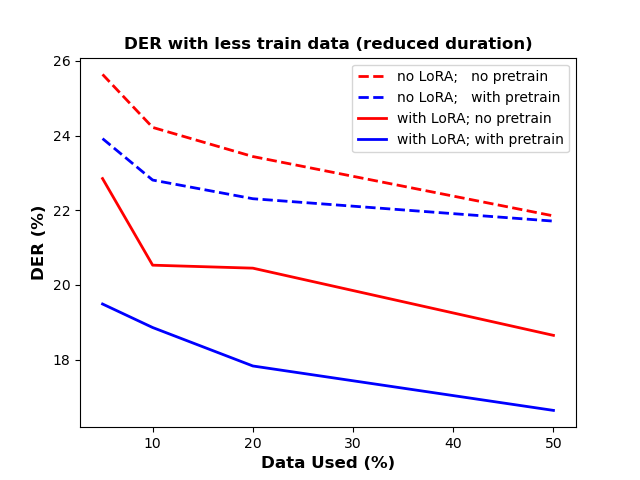}
  \label{fig:transfer_learning}
  \vspace{-5mm}
  \centering
  \includegraphics[width=8cm, height=5.5cm]{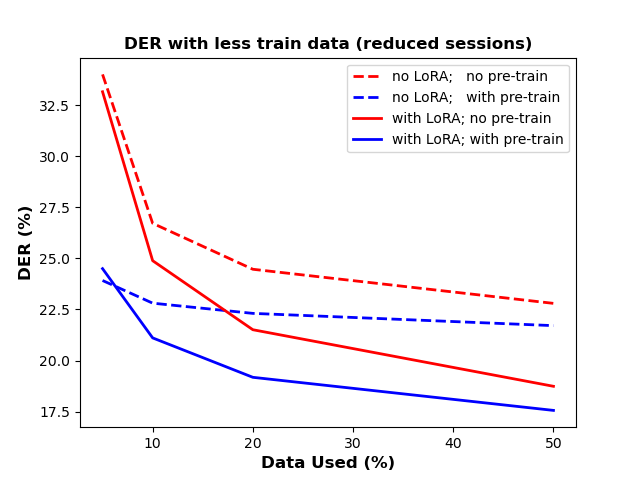}
  \vspace{-3mm}
  \caption{Transfer learning with pruned training data. With or without LoRA, and with or without pre-training on simulated conversations.}
  \vspace{-3mm}
\end{figure}

\subsection{Transfer Learning with Low Training Resources}
We prune the training data both at the session level, by excluding entire recording sessions, and at the duration level, by retaining only a predetermined fraction of each 15-minute audio clip. The results are shown in Fig~\ref{fig:transfer_learning}. As can be seen, pre-training on simulated conversations gives substantial advantages when fine-tuned on a very small amount of data. Even with only 5\% of data, which equals approximately 30 minutes of annotated data, the resulting DERs are lower than the DERs from the fully fine-tuned baseline speaker diarization systems. We also notice that adding LoRA reduces DER, especially when more train data are available for fine-tuning. 

\section{conclusion}
In conclusion, this work addresses the significant and unique challenge of limited publicly available child-adult speaker diarization models and annotated datasets. Specifically, we propose a data-efficient child-adult speaker diarization model framework using simulated child-adult conversations. We demonstrate competitive zero-shot speaker diarization performance by leveraging a pre-trained speech model, Whisper Encoder. Moreover, our findings indicate that fine-tuning on small amounts of real child-adult data further enhances speaker diarization performance, with LoRA further enhancing the transfer learning capabilities. Finally, we have made our code and model publicly available to facilitate advancements in automated child speech analysis for critical child-centered applications.

\bibliographystyle{IEEEtran}
\bibliography{refs}

\end{document}